\documentclass[11pt, oneside]{amsart}
\usepackage{geometry}                
\usepackage{graphicx}
\usepackage{amssymb}
\usepackage{epstopdf}
\usepackage{color}
\usepackage[nofiglist,nomarkers]{endfloat} %
\title[S.C. Kehr, Y.M. Liu, et al. \hfill Nat. Commun. 2:249 doi: 10.1038/ncomms1249 (2011) \hfill \hspace{0.2cm}]{Near-field examination of perovskite-based superlenses and superlens-enhanced probe-object coupling}

\begin{document}
\pagestyle{headings}
\maketitle

\begin{authors}
\begin{center}
{\bf S.~C.~Kehr}$^{1,2,\dagger,^*}$, {\bf Y.~M.~Liu}$^{3,\dagger}$, {\bf L.~W.~Martin}$^{1,4}$, {\bf P.~Yu}$^{5}$, {\bf M. Gajek}$^{5}$, {\bf S.-Y. Yang}$^{5}$, {\bf C.-H. Yang}$^{5,6}$, {\bf M. T. Wenzel}$^{7}$, {\bf R. Jacob}$^{8}$, {\bf H.-G. von Ribbeck}$^{7}$, {\bf M. Helm}$^{8}$, {\bf X. Zhang}$^{1,3}$, {\bf L. M. Eng}$^{7}$, {\bf and R. Ramesh}$^{1,5}$\\
\medskip
$^{1}$Materials Sciences Division, Lawrence Berkeley National Laboratory, Berkeley, CA 94720, USA\\
$^{2}$School of Physics and Astronomy, University of St Andrews, St Andrews, Fife KY16 9SS, UK\\
$^{3}$NSF Nanoscale Science and Engineering Center, University of California Berkeley, Berkeley, CA 94720, USA\\
$^{4}$Department of Materials Science and Engineering, University of Illinois, Urbana-Champaign, Urbana, IL 61801, USA\\
$^{5}$Department of Physics, University of California Berkeley, Berkeley, CA 94720, USA\\
$^{6}$Department of Physics and Institute for the NanoCentury, Korea Advanced Institute of Science and Technology (KAIST), Daejeon 305-701, Republic of Korea\\
$^{7}$Institute of Applied Physics, Technische Universit\"{a}t Dresden, 01062 Dresden, Germany\\
$^{8}$Institute of Ion Beam Physics and Materials Research, Helmholtz-Zentrum Dresden-Rossendorf, 01314 Dresden, Germany\\
$^{\dagger}$These authors contributed equally to this work.\\
$^{*}$ To whom correspondence should be addressed (sck21@st-andrews.ac.uk)
\end{center}
\end{authors}


\begin{abstract}
A planar slab of negative index material works as a superlens with sub-diffraction-limited imaging resolution, since propagating waves are focused and, moreover, evanescent waves are reconstructed in the image plane. Here, we demonstrate a superlens for electric evanescent fields with low losses using perovskites in the mid-infrared regime. The combination of near-field microscopy with a tunable free-electron laser allows us to address precisely the polariton modes, which are critical for super-resolution imaging. We spectrally study the lateral and vertical distributions of evanescent waves around the image plane of such a lens, and achieve imaging resolution of $\lambda/14$ at the superlensing wavelength. Interestingly, at certain distances between the probe and sample surface, we observe a maximum of these evanescent fields. Comparisons with numerical simulations indicate that this maximum originates from an enhanced coupling between probe and object, which might be applicable for multifunctional circuits, infrared spectroscopy, and thermal sensors.  
\end{abstract}

\newpage
\thispagestyle{headings}
\section*{Introduction}
In 1968 Veselago theoretically discussed a material with a negative refractive index, as well as its extraordinary responses to electromagnetic waves including negative refraction, reversed Doppler and Cerenkov effects and focusing with a planar lens \cite{Ves68}. Pendry revisited this idea 30 years later, showing that with such a planar lens one can create an image with a resolution beyond the conventional diffraction limit \cite{Pen00}. This opened the rapidly growing field of metamaterials, which present opportunities for new and remarkable applications in electromagnetics \cite{Sou07, Liu11} ranging from electrostatics \cite{Woo07, Mag08} via radio frequency \cite{Wil01}, microwave \cite{Smi00}, and Terahertz \cite{Yen04} to the infrared \cite{Lin04} and optical regime  \cite{Enk05, Lez07, Val08, Yao08}. The experimental proofs of negative refraction and sub-diffraction-limited resolution by a negative-index planar lens were first reported in the GHz range \cite{She01, Grb04}. Subsequently, it was shown for visible \cite{Fan05} and infrared (IR) \cite{Tau06} wavelengths that the evanescent field information of transverse-magnetic (TM) waves can be recovered using a medium with negative permittivity ($\varepsilon$) only. Such a lens is normally called a superlens, and substantial progress has been achieved in this rapidly developing area \cite{Zha08}.

A superlens is preferably realized by low-loss materials \cite{Pen00, Zha08}, among which perovskite oxides are good candidates. In the mid-infrared (mid-IR) range, perovskite oxides show phonon resonances, which are located at slightly different wavelengths for related materials such as e.g. bismuth ferrite (BiFeO$_3$) and strontium titanate (SrTiO$_3$) \cite{Spi62, Kam07}. On the high-frequency side of these phonon resonances, perovskites exhibit negative permittivities, which are suitable for superlens structures \cite{Pen00, Zha08}. At these wavelengths, the intrinsic absorption of light is small \cite{Spi62, Kam07}. In addition, epitaxially grown oxides exhibit highly crystalline interfaces resulting in low scattering. Both the small intrinsic and low scattering losses of perovskites could significantly improve the imaging resolution of superlenses. 

Perovskite oxides exhibit many intriguing properties such as colossal magnetoresistance \cite{Jin94}, ferroelectricity \cite{Ahn04}, superconductivity \cite{Wu87}, and spin-dependent transport \cite{Coe99}, which lead to numerous applications such as non-volatile memories \cite{Sco00}, microsensors and microactuators \cite{Set06}, as well as in nanoelectronics \cite{Was03}. Piezoelectricity and ferroelectricity allow for the manipulation of geometrical dimension, electric polarization, and dielectric properties by external electric fields \cite{Ram07}. Multiferroic perovskites, such as BiFeO$_3$, exhibit ferroelectricity as well as magnetoelectric coupling \cite{Ram07}, thus providing a pathway for additional degrees of tunability. Moreover, perovskites with matching lattice constants can be grown epitaxially on top of each other \cite{Schlom07, Schlom08}, which allows one to combine their properties in multifunctional heterostructures \cite{Schlom08, Mat97, Mar08}. These unique properties of perovskites may inspire new functionalities of metamaterial structures such as tunability by external fields and novel designs for multifunctional circuits.

In this paper, we study a new type of superlens for electric fields in the mid-IR based on the perovskite oxides BiFeO$_3$ and SrTiO$_3$.We investigate the evanescent waves in the image plane of perovskite superlenses by means of scattering-type near-field infrared microscopy (s-NSIM) \cite{Bet92, Zen95}. The combination with a free-electron laser (FEL), which is precisely tunable in the wavelength ($\lambda$) regime from 4 to 250~$\mu$m, enables us to address the polariton modes and study their lateral, vertical and spectral distribution. Such detailed characterizations are important to understand superlensing effect, but they were not comprehensively conducted in previous work \cite{Fan05, Tau06}. At certain wavelengths we observe enhanced evanescent fields in the image plane as well as a resolution beyond the classical diffraction limit. Moreover, we find that the evanescent fields show maxima at a certain distance between the probe and the sample surface, for which the vertical position depends on the wavelength. Comparisons with two-dimensional numerical simulations indicate that a superlens-enhanced coupling between probe and object causes this effect, which might be utilizable for controlling polariton propagation.

\section*{Results}
\subsection*{Perovskite-based superlens}\label{Experimental}

The original superlens proposed by Pendry is a single slab of a material with $\varepsilon=-1$ surrounded by air ($\varepsilon_{air}=+1$) \cite{Pen00}. At the two interfaces of the layer coupled surface polariton modes are excited, which amplify the evanescent electric fields arising from an object and transform them to the opposite side of the lens \cite{Pen00}. The oxide-based superlenses in our study consist of layers of matched perovskites (see Fig.~1a).  We structure a 50~nm thick film of metallic perovskite strontium ruthenate (SrRuO$_3$) on a SrTiO$_3$ substrate by photo-lithography, acting as superlens objects with a lateral size down to 3~$\mu$m. In order to ensure a fixed distance between objects and the slab we add a layer A (thickness $d$, $\varepsilon_A\cong\varepsilon_{air}=+1$) on top of the objects before we grow the actual superlens layer with $\varepsilon_B\cong-1$ (thickness 2$d$). Since layer A has a slightly different permittivity than air this two-layer superlens is asymmetric. Hence, we additionally study a superlens with symmetric design after adding an additional layer of material A (thickness $d$) on top. Superlensing is expected for such systems, when the real parts of the permittivities $\Re e(\varepsilon_A)$ and $\Re e(\varepsilon_B)$ have the same absolute values and opposite signs $\Re e(\varepsilon_B)=-\Re e(\varepsilon_A)<0$. Assuming the dielectric constants of the layers are similar to the ones described in literature \cite{Spi62, Kam07} (see Supplementary Fig. S1), this superlensing condition is fulfilled at $\lambda=13.9~\mu$m with the layers A and B being BiFeO$_3$ and SrTiO$_3$, respectively (Fig.~1b,c). Many other similarly suitable pairs of perovskites, such as PbZr$_{x}$Ti$_{1-x}$O$_3$ and SrTiO$_3$, BiFeO$_3$ and PbZr$_{x}$Ti$_{1-x}$O$_3$, BaTiO$_3$ and SrTiO$_3$, or BiFeO$_3$ and BaTiO$_3$, can be found at different wavelengths, implying the flexibility of perovskite superlenses in terms of the operation wavelength. Please note that other groups of dielectrics such as e.g. fluorides \cite{Axe65, Ram90} and simple oxides \cite{Bor93} can form similar pairs for superlenses, which altogether cover an even larger wavelength range of operation.

We examine the enhanced evanescent fields on the image side of the lens with s-NSIM in combination with a free electron laser (FEL) (see Fig.~1a and Methods), which allows for polariton-enhanced s-NSIM \cite{Keh08, Sch07}. Surface polariton modes can be excited at the interface of two materials which have permittivities of opposite signs \cite{Dio05, Rae88}.  In the studied wavelength range from 13.5$~\mu$m to 18.5~$\mu$m, the layers A and B of the perovskite-based superlenses show the following permittivities (see Figs.~1b,c) and corresponding polariton modes: When $\lambda>16.5~\mu$m we find $\Re(\varepsilon_B)<\Re(\varepsilon_A)<0$ and polariton modes are excited at the entire sample surface, being the interface between the toplayer and air. For $\lambda<16.5~\mu$m it is $\Re(\varepsilon_B)<0<\Re(\varepsilon_A)$ and polariton modes are created at both interfaces of layer B. As layer B is only 400 nm thick, these two modes are coupled, but are not necessarily localized.We observe an enhanced near-field signal, whenever a polariton mode is present at the position of the probe. However, only around the superlensing wavelength ($\varepsilon_A\cong-\varepsilon_B$,) the polartion modes are strongly localized and, hence, create a sub-diffraction limited image of the objects on the opposite side of the lens. Please note that with \lq near-field image\rq , we refer to the image obtained by near-field microscopy. The contrast in these images arises from local sample properties and thus different coupling to the probe, allowing us to distinguish the different objects on the opposite side of the superlens with a sub-diffraction-limited resolution (see Methods) \cite{Nov06}.

\subsection*{Mid-IR near-field imaging}
  
When placing the probe at a distance of about 30~nm to the sample, which is scanned relatively to the probe position, we obtain the near-field response as well as the topography of the sample. The corresponding results are depicted in Fig.~2 for similar object distributions with lateral sizes from 3x3$~\mu$m$^2$ to 8x8$~\mu$m$^2$ and three different samples, namely (a) structured SrRuO$_3$ objects on a SrTiO$_3$ substrate, (b) a symmetric superlens and (c) an asymmetric superlens without the top BiFeO$_3$ layer.  

The SrRuO$_3$ objects without a lens (Fig.~2a) show a near-field response, which depends on the polarization and the incident angle of the impinging light with respect to the objects. The $k$-vector direction of the incident TM-polarized light is shown in the sketch and in the topography image in (a) and is the same for all scans depicted in Fig.~2. Over a wide wavelength range in the IR, the permittivity of SrRuO$_3$ is much smaller than zero, resulting in an excellent, metal-like scattering behavior. Consequently, the distance dependence of the near-field signal on SrRuO$_3$ exhibits the same exponentially decaying character of the evanescent waves for all wavelengths (see middle panel of Fig.~2a).  The metallic SrRuO$_3$ structures show enhanced field intensities when their geometries match the wavelength, an effect which is well known in plasmonics \cite{Mue05}. Consequently, certain parts of the objects appear bright in the second- and third-harmonic near-field signals $NF_{2\Omega}$ and $NF_{3\Omega}$ (see Methods).  In general $NF_{2\Omega}$ is about 3 times larger than $NF_{3\Omega}$, with an interference-like background signal. In order to ensure pure near-field detection in the following, we will depict third-harmonic signals only.

For the symmetric superlens (Fig.~2b), at $\lambda=17.5~\mu m$ we observe a near-field signal due to a non-localized polariton mode at the sample surface, showing no clear contrast between areas with and without SrRuO$_3$ objects. Close to the superlensing wavelength, at around 14.6$~\mu$m, the evanescent fields are highly localized, resulting in a strong contrast reconstructing the SrRuO$_3$ structures. Please note that even though the NSIM probe scans more than 800~nm higher above the SrRuO$_3$ objects compared with the bare SrRuO$_3$ case, the signal is about 2 times stronger because the coupled polariton modes at the interfaces enhance the near-field signals arising from the objects. 

On the asymmetric superlens (Fig.~2c) we observe a similar, slightly blue-shifted response compared to the symmetric superlens, with non-localized polariton modes around 17.3$~\mu$m and superlensing with maximum contrast at around 14.1$~\mu$m. The signal on the asymmetric superlens is 2 times stronger than for the symmetric superlens, as the near-field probe is placed by 200~nm closer to the top SrTiO$_3$ surface at which the polariton mode is excited. 

With both superlenses we  clearly resolve the buried SrRuO$_3$ objects at the corresponding superlensing wavelengths. In the following, we discuss in detail the spectral response of both samples over a broad wavelength range.

\subsection*{Spectral response of the near-field signals}

Our s-SNIM setup allows for studying polariton modes with different characteristics in the wavelength range from 13.5$~\mu$m to 18.4~$\mu$m. 
For the symmetric superlens (Fig.~3), we observe an enhanced localized signal for $\lambda$ from 14 to 15.5~$\mu$m, close to the superlensing condition. The highest contrast is observed at 14.6~$\mu$m and the signal drops to zero for wavelength 1~$\mu$m smaller or larger than that. Note that areas without objects show a small negative signal in the superlensing regime, possibly due to evanescent waves on the substrate superlens interface with opposite phase. At longer wavelengths both layers, BiFeO$_3$ and SrTiO$_3$, show negative permittivities and a single polariton mode is excited at the topmost sample surface leading to an enhanced near-field signal on the entire sample area for $\lambda>17~\mu$m. Even though this signal is up to 8 times higher than the superlensed signal, it does not show any image contrast except for some artifacts at the topography edges caused by scattering and errors in the distance control. 

For the asymmetric superlenses (Fig.~4),
 we observe a localized near-field signal around the superlensing wavelength for 13.8$~\mu$m$~<\lambda<15~\mu$m with a clear object-related contrast at 14.1$~\mu$m.  In addition, the top-most layer SrTiO$_3$ of this lens supports a non-localized polariton mode for 14~$\mu$m$~<\lambda<15.5~\mu$m. In contrast to the symmetric case, in which this mode is spectrally well separated from the superlensing wavelength, for the asymmetric lens the wavelength regimes of superlensing and non-localized polariton overlap. Interestingly both modes respond with opposite phases resulting in a destructive superposition on the objects, which appear dark in the range from 14.5~$\mu$m to 15.2$~\mu$m. With increasing wavelength the non-localized mode dominates and the images are blurred.

 The smallest structures resolved on the superlenses are about 3x3~$\mu$m$^2$ in size and about 1$~\mu$m apart corresponding to a relative resolution of about $\lambda/14$. This sub-diffraction limited image is created by the superlensing effect. Compared to the image of bare SrRuO$_3$ objects the superlensed image appears homogeneously bright as the incident light is scattered by the probe and from there directed towards the object (see Methods). In general, the imaging with s-NSIM is influenced by the probe and its vertical position as it is in particular observed in the cross sections of Fig.~5, which will be discussed in the following paragraphs. 

\subsection*{Mid-IR near-field cross sections and spectroscopy}

We focus on the asymmetric superlens to study the vertical distributions of evanescent waves. Similar measurements for the symmetric superlens can be found in the Supplementary Fig.~S2. Fig.~5a depicts normalized vertical cross sections on the image side of the sample for different wavelengths showing the near-field signal as a function of the probe-sample distance $z$ and the position of the sample. The topography of the sample is reflected by the dark area on the bottom of the pictures with a 4$~\mu$m-wide SrRuO$_3$ object in the center.  For  $\lambda\geq14.8~\mu$m the probe excites propagating polariton modes close to the SrTiO$_3$ surface. Around the superlensing wavelength, for $\lambda=14.5~$to$~13.5~\mu$m, the evanescent field is localized on the object with decreasing signal and contrast for shorter wavelengths.
For the latter wavelength regime, we observe an intriguing phenomenon: a maximum in the evanescent field appears at a certain distance $z_0$ between tip and sample surface. Moreover, $z_0$ increases with smaller wavelength and appears at distances of up to 150~nm from the sample surface at $\lambda=13.5~\mu$m.  At first glance, this effect seems to be unexpected because the phonon polariton mode is a confined surface mode, which exponentially decays from the interface. 
To make sure that this effect is not an artifact in our experimental setup, we compare these results with numerical simulations as discussed in the following.

Fig.~5b shows the simulated data for a superlens consisting of the same constituents and geometry as in our experiments (see Methods). In these simulations we clearly observe the same effect as in Fig.~5a, that is, the field maximum gradually shifts away from the sample surface when the wavelength decreases. This maximum has an asymmetric shape, locating on the right-hand side of the structure, which is illuminated from the left with an incident angle of 75$^{\circ}$. The asymmetry might be formed by shadowing of the structure by the probe: when the probe is placed on the left side of the structure, it reflects the incident light and light hardly reaches the full structure; whereas when the probe is positioned on the right, the structure is completely illuminated by the beam. Compared to the experimental results, the simulations show some differences: Firstly, the wavelengths as well as the $z$-position of the maximum are slightly shifted. This is likely due to the small discrepancy between the dielectric constants of the fabricated superlens layers and the data reported in the published literature \cite{Spi62, Kam07}, which are used for the simulations (see Supplementary Fig.~S1). Secondly, the near-field signal above the topographic step for $\lambda\geq14.5~\mu$m in the simulation shows a saddle shape. The enhanced signal at the topographic steps is possibly caused by the sharp edges in the modeled geometry, which give rise to a highly nonlinear near-field due to localized corner polariton modes \cite{Ber00}. Please note that this effect is only relevant in the propagating polariton regime, whereas it is much less pronounced around the superlensing wavelengths since the polariton mode at the flat interface of the step becomes also highly localized.  In the supplementary information (Supplementary Fig.~S3), we compare these results with corresponding simulations without the topographic step. We find very similar spectral, lateral and vertical response, but the enhanced signal at the edges disappears. This comparison verifies that the small topography on the sample interfaces due to the sample fabrication has negligible influence on the observed optical signals. The enhanced coupling of probe and object by the superlens layer is indeed correlated to the material property rather than the topological protrusion.

We further compare the results of experiment (Fig.~5c) and simulation (Fig.~5d) by analyzing the absolute values of the maximum near-field signals with and without SrRuO$_3$ object, $NF_{w}$ and $NF_{w/o}$, respectively, as well as the resulting contrast $V=(NF_{w}-NF_{w/o})/(NF_{w}+NF_{w/o})$. Without object (green curves) we observe maxima at 15.3~$\mu$m (c) and 14.5~$\mu$m (d) due to non-localized polariton modes, whereas on the SrRuO$_3$ objects (red) an enhanced signal is observed for slightly shorter $\lambda$. The resulting contrasts (yellow) show maxima at 14~$\mu$m (c), and 13.3~$\mu$m (d), respectively, which is in good agreement with the predicted superlensing wavelength of 13.9$~\mu$m.

\subsection*{Numerical simulations of superlens-enhanced coupling}

Why do we observe an enhanced signal at a certain distance to the sample surface? The coupled phonon-polariton modes, which create the sub-diffraction-limited image, result in a field which decreases exponentially with the distance.  However, as we place a scattering probe on the image side, this probe itself acts as an object next to the superlens creating additional fields at the position of the SrRuO$_3$ objects. The probe-sample system therefore consists of two coupled scatterers with a superlens structure between both of them, showing a resonance as a function of the wavelength and the probe-sample distance as discussed in the following paragraphs.

Fig.~6a plots the electric-field distribution of an asymmetric BiFeO$_3$-SrTiO$_3$ superlens with a line-source in the object plane for different wavelengths and no probe on the image side. For wavelengths larger and smaller than the superlensing wavelength around $\lambda=13.5~\mu$m, we observe unconfined evanescent waves on the image side of the superlens due to the excitation of non-localized surface polariton modes at the SrTiO$_3$-air interface. For $\lambda=13.5~$and$~14~\mu$m, the superlensing effect takes place with an enhanced confined field on the image side of the lens and localized polariton modes at both interfaces of the SrTiO$_3$ layer. The superlensing effect is further verified by the transfer function simulation of electric fields. Figure 6b shows the isothermal contour of the transfer function of the asymmetric superlens (for the transfer function of the symmetric superlens see Supplementary Fig.~S4). Around 13.5~$\mu$m, the amplitude of TM waves at the top interface of the superlens still maintains reasonably large, even for a tangential wave vector up to 10$k_0$ (Fig.~6c). In contrast, the field intensity dramatically decays for large wave vectors in a control sample, in which the 400~nm SrTiO$_3$ film is replaced by a 400~nm BiFeO$_3$ layer. 

Figure 6a shows that the electric field at the SrTiO$_3$-air interface possesses in all cases a maximum at the sample surface, reflecting the exponentially decaying behavior of the polariton mode. The situation changes with the presence of a scattering probe on the imaging side of the sample. Fig.~6d shows a simulation for the ideal case of two equally sized metal spheres, object and probe, separated by the 2-layer superlens for a fixed wavelength of 14~$\mu$m and for different gaps $z$ between the top scatterer (i.e., the probe) and the sample surface. For small gaps $z$=25~nm, one observes a large field underneath the probe-scatterer $E_{tip}$ due to near-field enhancement as well as a localized polariton mode $E_{int}$  at the interface between SrTiO$_3$ and BiFeO$_3$. When the gap is increased to 75~nm,  $E_{tip}$ decreases, but $E_{int}$ is much larger than that for $z$=25~nm, whereas both fields decrease with $z$ for gaps larger than 100~nm.

In Fig.~6e, we plot the $z$-dependence of $E_{tip}$ (green) and $E_{int}$ (red), as well as of the integrated Poynting vector far away from the two objects $S$ (yellow) that corresponds to the scattered light intensity measured in our experiments. $E_{tip}$ decreases exponentially with the distance, because of the exponential decay of polariton modes at the sample surface. However,  $E_{int}$ and $S$ show maxima at certain distances between tip and sample of 70~nm and 50~nm, respectively. 
The polariton modes at the interfaces of the superlens enhance the evanescent waves arising from the object. Hence, a larger field amplitude of polariton modes $E_{int}$ for a certain probe-sample distance indicates enhanced superlensing of the coupled probe-object system for this position of the probe scatterer. Moreover, the scattered light $S$ from this system has a maximum around the same gap-size, which correlates the far-field observation with the localized enhanced coupling.

\section*{Discussion}

The theory of transformation optics \cite{Leo06,  Leo06b, Pen06} states that a superlens with negative index of refraction distorts the optical space in a remarkable way: the space is folded by the superlens, with the object plane and the image plane (as well as a plane within the lens) at the same position in optical space \cite{Leo06}. If we place scatterers in both planes, one can consider them to be two induced dipoles located at exactly the same position in optical space, which therefore might lead to the observed enhanced coupling of both scatterers. 

We note that a maximum in the image plane of a superlens was also observed earlier in the microwave range using loop antennas as source and detector \cite{Guv04, Mes05, Fre05}. The position of the maximum changed due to matched coupling between object and probe, when using detector antennas with different radii and different resistance load \cite{Mes05}. However, any dependences on the wavelength and on corresponding changes in the sample dielectric constants were not studied. It was proposed that this effect could be applicable for three-dimensional imaging \cite{Mes05}, which might be possible with the oxide-based superlens as well.

In this paper, we proposed, designed and demonstrated perovskite-based superlenses for electric fields. These materials are in particular suitable for superlenses in the infrared range, showing matching pairs of the real parts of their dielectric constants with opposite signs. Moreover, perovskites show low intrinsic absorption at the wavelengths of interest and different materials can be grown epitaxially on top of each other. Both effects lead to low losses in the superlens structure. As some perovskite oxides are ferroelectric (e.g. BiFeO$_3$ in the current study), a superlens consisting of these materials might be tunable by an external electric field.

We characterized the perovskite superlenses in both spectral and three-dimensional spatial domain. The near-field examination shows sub-wavelength resolution of $\lambda/14$ at the superlensing wavelength. Comparison of symmetric and asymmetric superlens show a stronger superlensed signal for the asymmetric case  in which the probe is placed directly at the superlens interface. This finding is supported by the calculated transfer function of the asymmetric superlens, which shows a better performance compared to the symmetric case in terms of field enhancement and wave vector bandwidth.

In addition, we discussed the  coupling of near-field probe and object, being placed on two opposite sides of a superlens. It is found that the coupling effect is strongly enhanced at phonon-polariton resonances and is dependent on the probe-object distance. These findings reflect the fact that in a system consisting of object, superlens, and detecting probe coupling between all constituents takes place. The observed superlens-mediated interaction between two particles might allow for controlled enhanced-coupling effects. These effects could find potential applications in local thermal sensors as described by Shen \textit{et al.} \cite{She09} as well as in metamaterial-based multifunctional circuits \cite{Eng05, Eng07}. We envision for example an enhanced transport of polaritons from particle to particle by these superlenses: a series of superlens-coupled particles can act as a polariton conductor with narrow bandwidth, which is determined by the superlensing material.   

\section*{Methods}

\subsection*{Growth and absorption losses of perovskite oxides}

The various thin films were grown via pulsed laser deposition. Thin films of SrRuO$_3$ were grown at 680$~^{\circ}$C and films of BiFeO$_3$ and SrTiO$_3$ were grown at 700$~^{\circ}$C in 100~mTorr of oxygen \cite{Chu07}. Following the growth of the SrRuO$_3$ layer, features were defined via photolithography and samples were ion-milled to produce the objects; subsequently, BiFeO$_3$ and SrTiO$_3$ films were grown. Films were found to be single phase and fully epitaxial in all cases via X-ray diffraction and transmission electron microscopy. 

In order to compare the absorption of our perovskite-based superlens with former materials used, we calculate the ratios $n/\kappa$ with $n$ and $\kappa$ being the real and imaginary parts of the refractive index at the wavelengths for which superlensing is expected. Larger $n/\kappa$ corresponds to less material absorption losses. For SrTiO$_3$ this ratio is 12.7$\cdot 10^{-2}$, three times larger than for SiC and silver, which both show a $n/\kappa$ of about 4.2$\cdot 10^{-2}$ at their corresponding superlensing wavelengths \cite{Tau06, Fan05}. 
Note that the superlensing condition for the perovskite-based superlenses is fulfilled for $\varepsilon_{STO} \cong-2$, whereas for the SiC and silver superlens it was observed around $\varepsilon \cong-3$. However, even when $\varepsilon_{STO} \cong-3$ at $\lambda=14.7~\mu m$, we find $n/\kappa=10.5\cdot 10^{-2}$ being much larger than for the other superlenses.

\subsection*{s-NSIM setup}

In s-NSIM, a scattering probe, namely a metal-coated atomic-force microscope (AFM) tip with a typical radius of 50~nm, is placed close to the sample surface \cite{Bet92, Zen95}. This probe transforms the evanescent fields into propagating waves, which can be detected in the far-field \cite{Bet92, Kno00}. In order to separate the near-field from the much larger far-field signal, we use the method of higher-harmonic demodulation \cite{Kno00, Wur98}: in tapping-mode AFM the distance between tip and sample is modulated with small amplitudes of about 30~nm at frequencies around $\Omega=150~kHz$. On this scale, the far-field changes linearly with distance resulting in a modulation with the same frequency $\Omega$. On the other hand, evanescent waves depend non-linearly on the distance, resulting in a modulation at $n\cdot \Omega$ ($n=1,2,3,...$). Hence, when filtering the modulated signal at $n\cdot \Omega$ with $n\geq2$, at higher harmonics, we obtain near-field components only, with less far-field contributions for larger orders of $n$. If not stated otherwise, the near-field signals shown in this paper represent third-harmonic signals $NF_{3\Omega}$ in order to ensure pure near-field examination. For comparisons at different wavelengths, we nomalize $NF_{3\Omega}$ to the current laser power and to the spectral response of detector and optical elements in the beam path.

Our investigation was performed using the free-electron laser FELBE at
the Helmholtz-Zentrum Dresden-Rossendorf (www.hzdr.de), Germany, which offers continuous
tunability across a wavelength range of $4-250~\mu$m at an average
power of up to 10~W (delivered as a picosecond pulse train at a
repetition rate of 13~MHz). In the 10 to 20~$\mu$m wavelength regime the typical spectral width of the FEL is $FWHM$=50-100~nm.

\subsection*{Near-field interaction and probe-sample coupling}

When imaging a superlensed signal by s-NSIM, the near-field probe can not be assumed to be a passive element only, but the coupling between probe and objects needs to be taken into account. The origin of the coupling lies in the optical interaction principle of NSIM, which results in the consequences for the NSIM examination of superlenses as discussed in the following. 

In our s-NSIM  and the similar one employed in Ref. \cite{Tau06}, the probe-sample system is illuminated from the probe side and, hence, the initial near field is generated by the probe rather than by the sample.
The function of the probe is two-fold: firstly, the evanescent waves arising from the probe illuminate the sample locally and excite modes in the sample and on its surface. Secondly, the probe senses the fields of the sample modes and transforms them by scattering into propagating waves, which can be detected in the far-field. This probe-sample interaction depends on the scattering behavior of the probe, the local optical properties of the sample, as well as the distance between the probe and the sample surface. As a first-order approximation, it can be described by the dipole model introduced by Knoll and Keilmann \cite{Kno00}: here, the s-NSIM signal is described by scattering in the near-field coupled probe-object system represented through an effective polarizability. A more sophisticated approach describes the probe-sample interaction through higher-order modes \cite{Ren05} showing the same qualitative results.

A critical point for successful NSIM measurements is that the probe properties and illumination are kept constant during the measurements, therefore the changes in the near-field signature are essentially caused by local sample properties. In our study we particularly keep the position of the probe fixed while the sample is scanned to ensure constant illumination. The probe properties like material, geometry and orientation with respect to the polarization of the incident light are fixed as well. Hence, the observed contrast arises from changes in the sample only. 
The influence of probe or object properties are enhanced, when either one of the two is excited close to its resonances. In the mid-IR a metallic probe shows no distinct resonances whereas the samples discussed in our work support polariton resonances. Hence, it is the sample that determines the NSIM signal, whereas the tip acts solely as a small optical dipole.

In the case of a superlens as the sample the near-field interaction is extended: at the superlensing wavelength, the evanescent waves arising from the probe are transferred to the object plane. The objects are excited by these waves and generate additional evanescent fields, which are reconstructed by the superlens on the image side of the sample. Finally, the near-field signal at the image plane is transformed by the probe into detectable propagating waves. As the observed signal depends strongly on the properties of the superlens as well as the positions of both scatterers, probe and object, with respect to the superlens slab, we call this effect a superlens-enhanced probe-object interaction. 

Besides the superlens-enhanced interaction between probe and objects, the probe can also interact with the surface layer of the superlens, which is in our case either BiFeO$_3$ or SrTiO$_3$ for the symmetric or asymmetric superlens, respectively. We observe enhanced near-field signals due to propagating polariton modes  at the toplayer-air interface of the samples at wavelengths around 17~$\mu$m and 14.6~$\mu$m for symmetric and asymmetric superlens, respectively. Please note that, unlike the localized modes at the superlensing wavelength, this signal does not carry any information about the SrRuO$_3$ objects (see Figs.~3 and 4).

\subsection*{Numerical simulations}

The numerical simulation results presented in the manuscript are all based on the commercial finite-element solver COMSOL 3.5. Due to memory limitation, the simulation is performed in two dimensions. The extremities of the simulation domain are assigned scattering properties which essentially mimic the necessary open boundary conditions. We modify the size of simulation domain and (local) meshes to ensure that the field variation is less than 1~\%. Such a convergence of the numerical simulation verifies that the boundary condition and meshing are assigned properly. The total mesh number is up to 450,000. For a workstation with 32G RAM memory and two dual CPUs (2.66~GHz), each simulation takes less than two minutes to converge and reach the relative tolerance of 10-6. 

To numerically retrieve the s-NSIM signal shown in Fig. 5, iterative simulations with varying tip positions are performed by combining a Matlab script with COMSOL.  The near-field probe is made by gold, whose dielectric constant is taken by fitting the data from Johnson and Christy \cite{Joh72}. The probe is modeled as a triangle, whose tip angle is about 22 degrees and tip apex is a half sphere with the diameter of 100~nm. The superlens consists of layers of BiFeO$_3$ and SrTiO$_3$ with thicknesses of 200$~$nm and 400~nm, respectively, and the dielectric constants are taken from literature \cite{Spi62, Kam07}. The SrRuO$_3$ objects are assumed to be rectangles of 50$~$nm$\times4~\mu$m with a dielectric constant of SrRuO$_3$ determined by Fourier transform infrared (FTIR) spectroscopy (see Supplementary Fig.~S1). The system is illuminated by monochromatic light with an incident angle of 75$^{\circ}$. In order to study the response of the system, we calculate the electric field distribution at all positions of the system. We observe a field enhancement underneath the tip and at the superlens interfaces due to polariton excitation. The scattering cross sections (Fig.~5b) as measured in the experiment are related to the integrated Poynting vector over a solid angle about 70$^{\circ}$ from the tip apex. Such a simulation is repeated for different tip positions, in direct analogy to the scanning process. In order to simulate the higher-harmonic demodulation, we calculate the gradient of the Poynting vector to obtain the first-harmonic signal. Consequently, the gradient of the first harmonic reflects the second-harmonic signal and so on \cite{Hil00}.

 \section*{Acknowledgements}

We like to thank G. Bartal, Y. Wang, X. Yin, L. Zeng, and J. Seidel for fruitful discussions. 
We acknowledge the technical assistance of the FELBE team at Helmholtz-Zentrum Dresden Rossendorf and thank them for their dedicated support.

The work at Berkeley was supported by the Director, Office of Science, Office of Basic Energy Sciences, Materials Sciences Division of the U.S. Department of Energy under contract No. DE-AC02-05CH11231. 
S.C.K was supported by the German Academic Exchange Service DAAD. 
Y.M.L and X. Z. acknowledges funding support from US Army Research Office (ARO) MURI program 50432-PH-MUR. M.T.W., R.J., H.-G.v.R., M.H., and L.M.E. gratefully acknowledge the funding by the German Science Foundation DFG (projects HE 3352/4-1 and EN 434/22-1). M.T.W. and H.-G.v.R. were supported by the EU-STREP Project in FP7 ÒPLAISIRÓ.

\section*{Author contributions}

S.C.K developed the concept, designed the samples and the experiments, carried out the NSIM measurements, analyzed the data, interpreted the experimental results, and wrote the manuscript. Y.M.L. performed all numerical simulations, carried out the FTIR measurements, interpreted the experimental results, and wrote the manuscript. L.W.M., P.Y., S.-Y.Y, and C.-H.Y grew the perovskite films. M.G. prepared the structures by means of photo-lithography. M.T.W., R.J. and H.-G.v.R. performed NSIM measurements. M.H. conducted the NSIM at the FEL facility. L.M.E. supervised the NSIM and developed the concept. X.Z. supervised the numerical simulations and Fourier transform infrared measurement and participated in the manuscript revisions. R.R. conducted the growth of all perovskites.

\section*{Additional information}
For Supplementary Information see http://www.nature.com/naturecommunications.\\
Competing financial interests: The authors declare no competing financial interests. \\
Reprints and permission information: http://npg.nature.com/reprintsandpermissions/ .\\
How to cite this article: Kehr, S.C. et al. Near-field examination of perovskite-based superlenses and superlens-enhanced probe-object coupling. Nat. Commun. 2:249 doi: 10.1038/ncomms1249 (2011).\\
License: This work is licensed under a Creative Commons Attribution-NonCommercial-Share Alike 3.0 Unported License. \\To view a copy of this license, visit http:// creativecommons.org/licenses/by-nc-sa/3.0/ .\\

 \newpage
 
\begin{figure}[htbp]
\begin{center}
\includegraphics[width=0.9\textwidth]{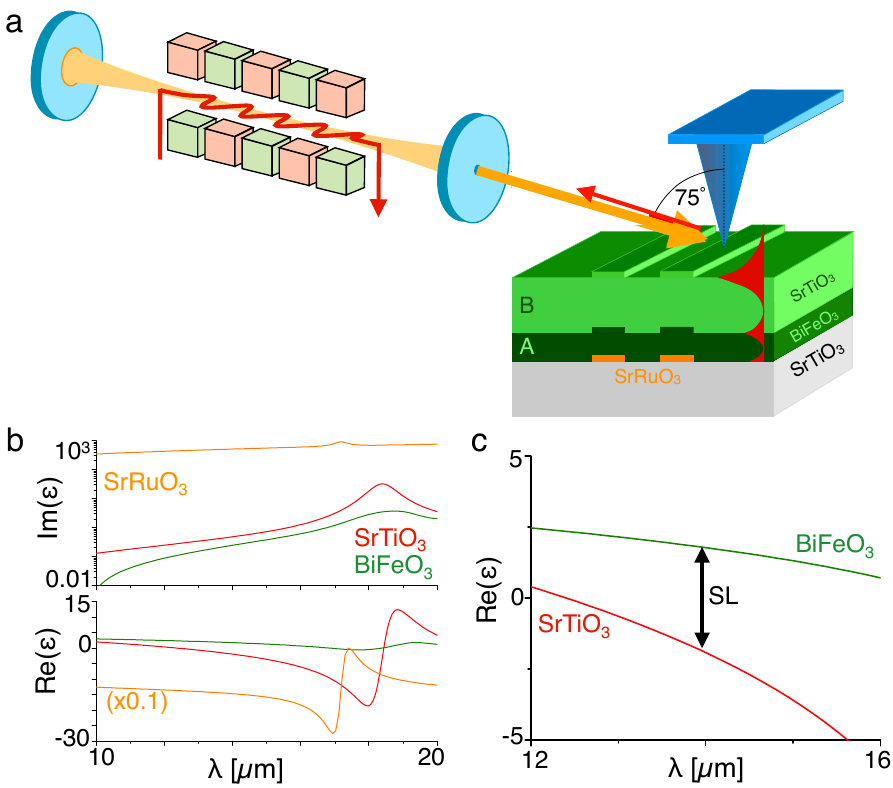}
\caption{\textbf{s-NSIM setup and perovskite properties} (a) Sketch of the experimental setup including the superlens and, the geometry at the near-field probe (blue), and the free-electron laser light source. The superlens consists of the layers A (BiFeO$_3$) and B (SrTiO$_3$) of thicknesses $d$ and 2$d$ ($d$=200~nm), respectively. The objects to be imaged are SrRuO$_3$ patterns on a SrTiO$_3$ substrate. All constituents of the superlens are perovskite oxides that match in their crystalline structure resulting in low scattering at the highly crystalline interfaces. The near-field tip probes the evanescent fields on the image side of the lens. The superlens is excited by an IR free-electron laser, which is precisely tunable in the range from 4 to 250~$\mu$m. (b) Imaginary and real parts of the dielectric constants $\varepsilon$ of all constituents (SrRuO$_3$ data determined by FTIR spectroscopy, values for SrTiO$_3$ and BiFeO$_3$ taken from literature \cite{Spi62, Kam07}, see Supplementary Fig.~S1); (c) real parts of the dielectric constants at the high-frequency side of their phonon resonances depicted in b. The arrow indicates the wavelength at which superlensing is expected.}
\end{center}
\end{figure}

\begin{figure}[htbp]
\begin{center}
\includegraphics[width=0.9\textwidth]{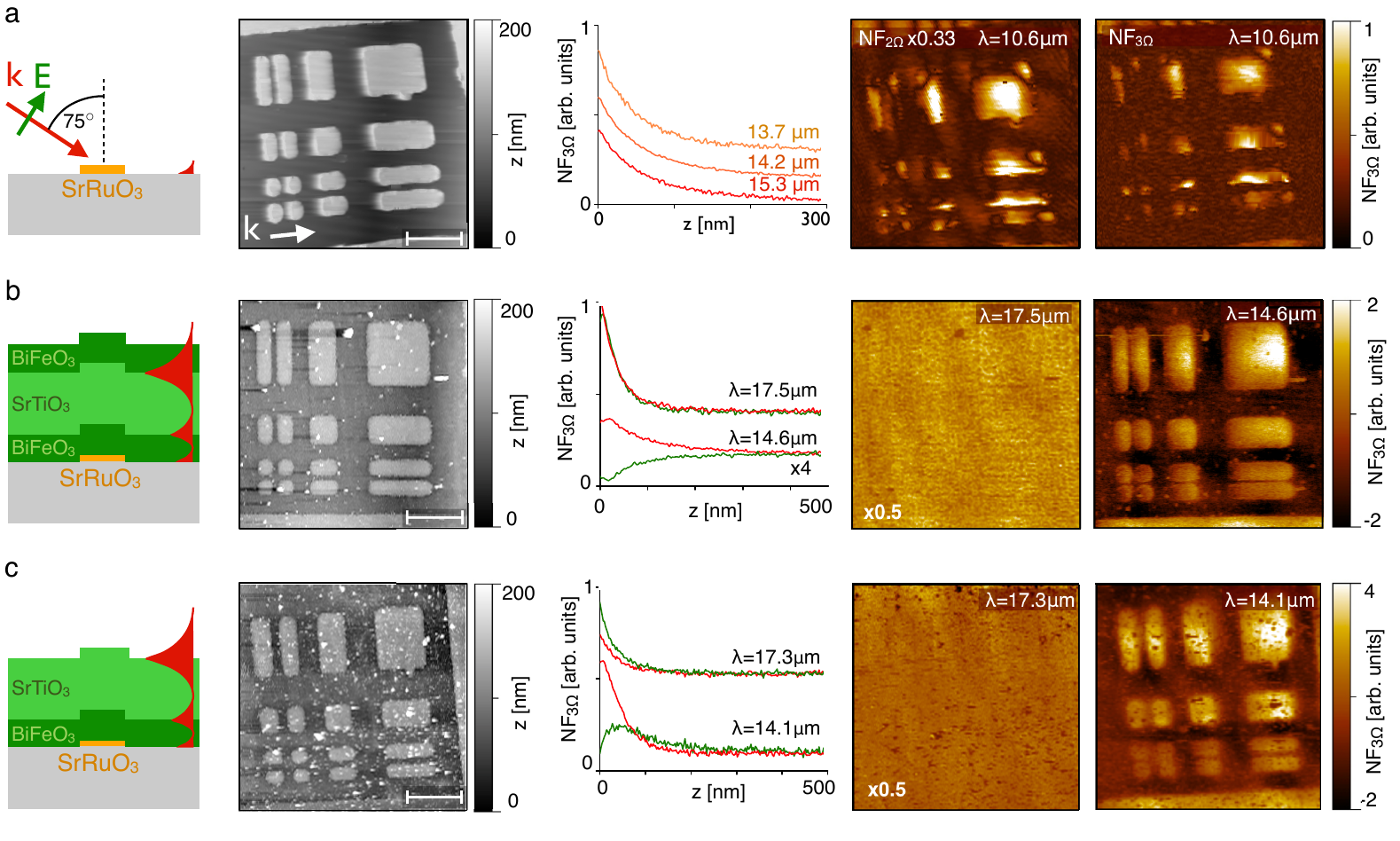}
\caption{\textbf{Near-field images of three different samples.} From left to right, the subfigures display: sketches of the sample, topography images obtained by atomic force microscopy (scalebars are 10$~\mu$m), near-field signals as functions of the probe sample distance $z$ for selected wavelengths (an offset is added for better comparison), as well as near-field images as described in the following. (a) For SrRuO$_3$ objects on a SrTiO$_3$ substrate we image second- and third-harmonic near-field signals ($NF_{2\Omega}$ and $NF_{3\Omega}$) using a CO$_2$ laser ($\lambda=10.6~\mu$m). (b), (c) For both types of superlenses we depict $NF_{3\Omega}$ at two different wavelengths being $\lambda=17.5~\mu$m and 14.6~$\mu$m for the symmetric superlens (b) and $\lambda=17.3~\mu$m and 14.1$~\mu$m for the asymmetric lens (c). The red and green curves in the distance curves correspond to areas with and without SrRuO$_3$ objects on the opposite side of the lens, respectively. Such distance curves and the near-field images show strong signals at both wavelengths, but only at the shorter wavelengths we observe a contrast beyond the diffraction limit  due to the superlensing effect.}
\end{center}
\end{figure}

\begin{figure}[htbp]
\begin{center}
\includegraphics[width=0.9\textwidth]{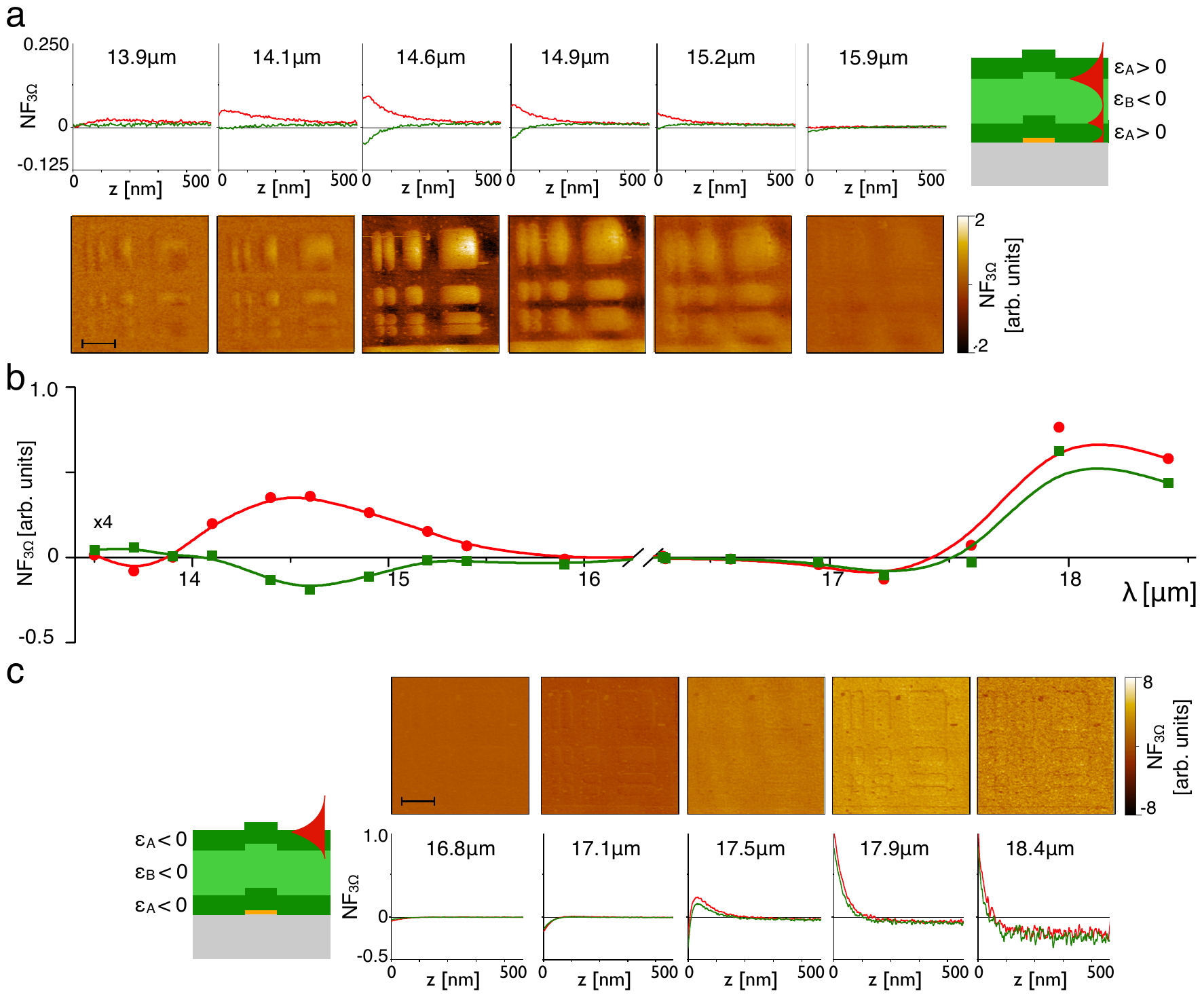}
\caption{\textbf{Spectral response of the symmetric superlens.} Near-field spectrum as well as near-field images and $NF_{3\Omega}$-distance curves for selected wavelengths (scalebars are 10$~\mu$m). (a) For $\lambda=13.9$~to~$15.9~\mu$m, an imaging contrast exists due to the localized polariton mode. (b) Shows the near-field spectrum for a fixed distance of $z$=20~nm with the results between 13.5~$\mu$m and 16.25~$\mu$m being multiplied by a factor of 4 in the plot. (c) For $\lambda=16.8$~to~$18.4~\mu$m,  no imaging contrast is observed although the near-field signal is enhanced by the non-localized polariton mode. The red and green curves in the distance curves of a,c and the spectrum (b) correspond to the near-field signals on areas with and without SrRuO$_3$ objects on the opposite side of the lens, respectively.}
\end{center}
\end{figure}

\begin{figure}[htbp]
\begin{center}
\includegraphics[width=0.9\textwidth]{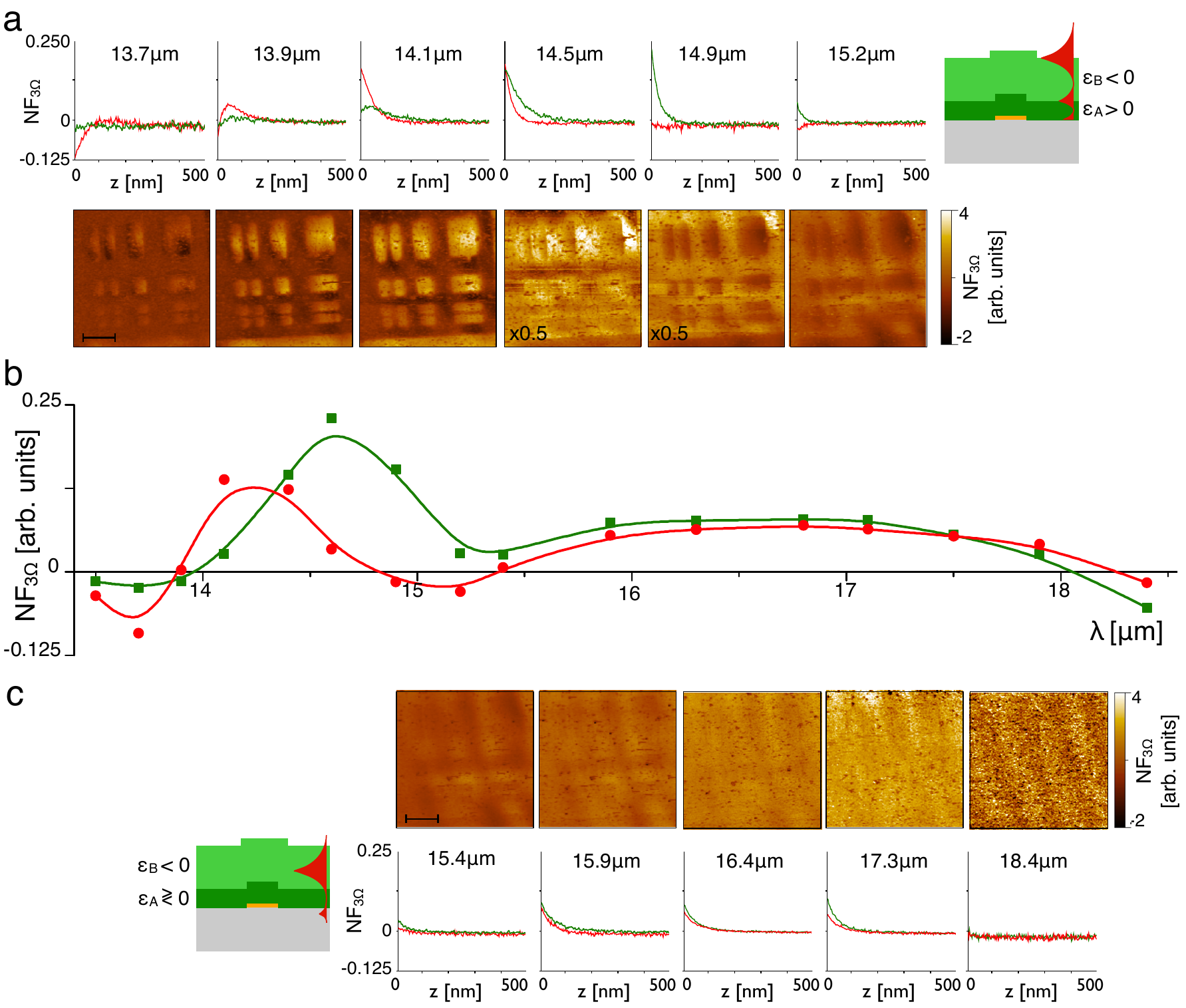}
\caption{\textbf{Spectral response of the asymmetric superlens.} Near-field spectrum as well as near-field images and $NF_{3\Omega}$-distance curves for selected wavelengths (scalebars are 10$~\mu$m). (a) For $\lambda=13.7$~to~$15.2~\mu$m, an imaging contrast exists due to the localized polariton mode. (b) Shows the near-field spectrum for a fixed distance of $z$=20~nm. (c) For $\lambda=15.4$~to~$18.4~\mu$m,  no imaging contrast is observed although the near-field signal is enhanced by the non-localized polariton mode. The red and green curves in the distance curves of a,c and the spectrum (b) correspond to the near-field signals on areas with and without SrRuO$_3$ objects on the opposite side of the lens, respectively.}
\end{center}
\end{figure}

\begin{figure}[htbp]
\begin{center}
\includegraphics[width=0.9\textwidth]{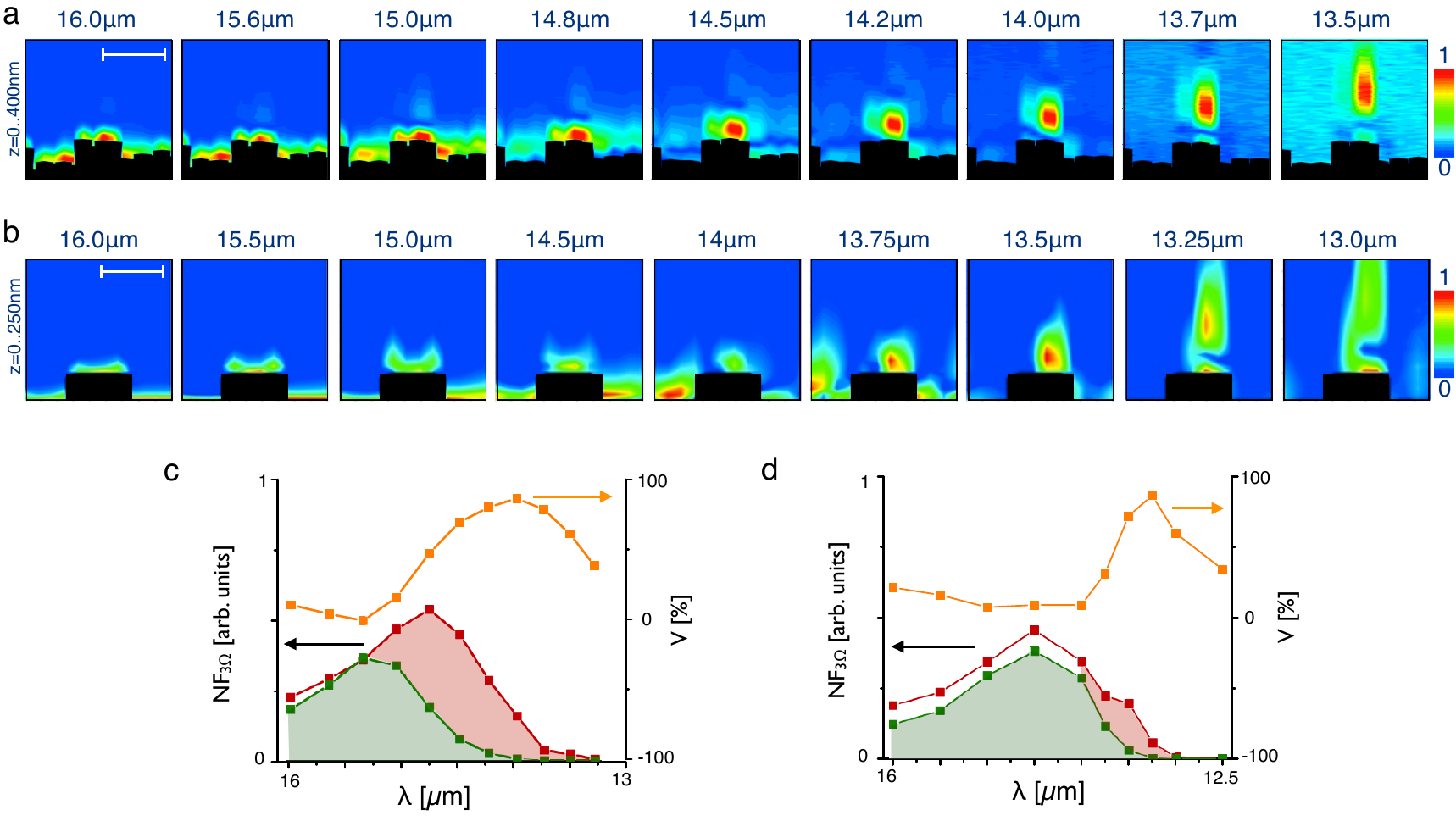}
\caption{\textbf{Normalized cross sections on an asymmetric superlens.} (a) depicts experimental data, whereas (b) shows the results from numerical simulations (for details see text).  The horizontal range in all cross sections is 12$~\mu$m (scalebars are 6$~\mu$m). The topography of the sample and the position of the SrRuO$_3$ object is reflected by the dark areas at the bottom of the figures. The localized evanescent fields on the objects show a maximum at a certain distance $z_0$, which increases with smaller wavelength. This effect is observed experimentally as well as in the simulations and corresponds to a superlens-enhanced coupling of probe and object. (c,d) Spectral behavior of the near-field maxima in the experiment (c) and in the simulations (d) for areas with (red) and without (green) objects as well as the corresponding contrast $V$ (yellow) calculated from this data. The marked areas correspond to the observation of non-localized near-field signals (green), and localized evanescent fields due to the superlensing effect (red).}
5\label{default}
\end{center}
\end{figure}

\begin{figure}[htbp]
\begin{center}
\includegraphics[width=0.9\textwidth]{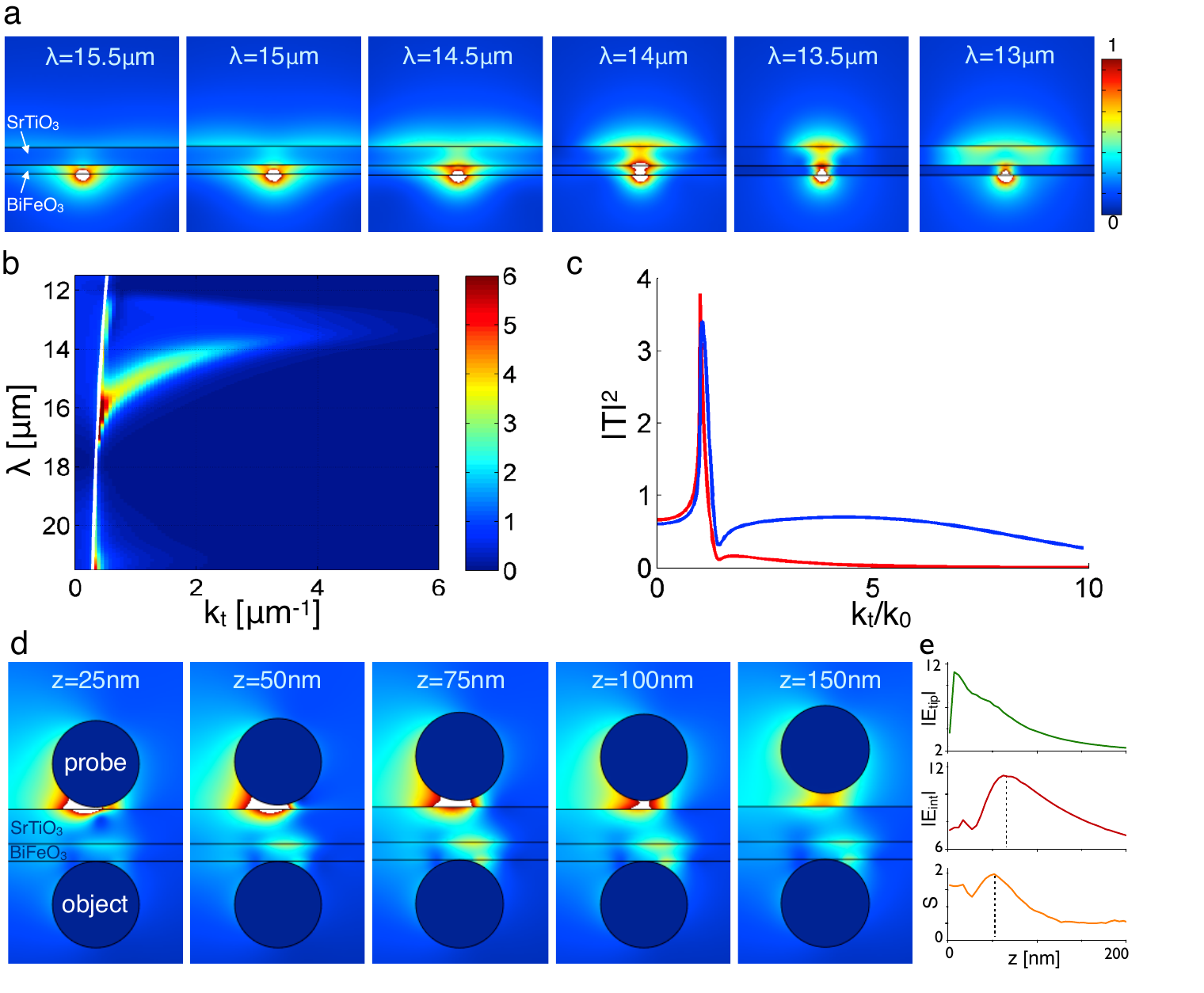}
\caption{\textbf{Simulations of electric field distributions and transfer function for the asymmetric superlens.} (a) Planar superlens with a line-source as object on one side for different wavelengths, showing a confined field on the image side of the lens for $\lambda=13.5-14~\mu m$ due to superlensing (all figures with same color scale in arbitrary units). (b) The isothermal contour of transfer function in the wavelength range of our interest plotted versus wavelength $\lambda$ and wavevector $k_t$. The color represents the transfer function (the square of the ratio between the transmitted electric field after the superlens and the incident field). The white line is the light line in air. (c) Transfer functions $|T|^2$ for the asymmetric superlens (blue) and the control sample (red) at 13.5~$\mu$m wavelength. The control sample replaces the 400~nm SrTiO$_3$ film in the superlens by a 400~nm BiFeO$_3$ layer. One can clearly see that the evanescent wave is enhanced by the superlens over a large range of wave vectors (up to 10$k_0$). The sharp peaks around $k_0$ are due to total internal reflection. (d) A planar superlens with two spherical objects on both sides for $\lambda=14~\mu m$ and increasing gap $z$ between the upper sphere (probe) and the sample surface. (e) Parameters of interest extracted from simulations as shown in d: the electric field at the lower apex of the probe sphere $E_{tip}$, the electric field at the SrTiO$_3$-BiFeO$_3$ interface $E_{int}$, and the integrated Poynting vector $S$ far away from the two-sphere system. All parameters are depicted as a function of the gap $z$. In contrast to $E_{tip}$, which has the highest value at $z\to0$, $E_{int}$ and $S$ show maxima for certain $z$ being 70~nm and 50~nm, respectively.}
\end{center}
\end{figure}

\end{document}